# Moment of inertia of a trapped superfluid gas of atomic Fermions

M. Farine[1], P. Schuck[2], X. Viñas[3]

[1]Ecole Navale, Lanvéoc-Poulmic, 29240 Brest-Naval, France

[2]Institut des Sciences Nucléaires, Université Joseph Fourier, CNRS-IN2P3

53, Avenue des Martyrs, F-38026 Grenoble Cedex, France

[3]Departament d' Estructura i Constituents de la Mat'eria Facultat de F'isica,

Universitat de Barcelona Diagonal 647, E-08028 Barcelona, Spain

**Abstract:** The moment of inertia $\Theta$ of a trapped superfluid gas of atomic Fermions ($^6$Li) is calculated as a function of the temperature. At zero temperature the moment of inertia takes on the irrotational flow value. Only for T very close to $T_c$ rigid rotation is attained. It is proposed that future measurements of the rotational energy will unambiguously reveal whether the system is in a superfluid state or not.

## 1. Introduction

The advent in 1995 of Bose-Einstein-Condensation (BEC) of atomic Bosons in magnetic traps certainly represents a milestone in the study of bosonic many body quantum systems. This is so because a systematic study of these systems, starting with the free particle case, as a function of increasing density, particle number, and other system parameters seems possible and has already progressed to a large extent while going on at a rapid pace [1,2]. On the other hand the recent experimental achievement of trapping $^6$Li atoms and other fermionic alkali atoms [3] also spurs the hope that for the fermionic many body problem, as much progress will be made in the near future as for the bosonic systems. Indeed the first Fermi-Dirac degeneracy of trapped $^{40}$K atoms has already been observed (see B. De Marco and D.S. Jin [4]). In this reference also more of the physics of trapped Fermionic atoms is discussed. For atoms with attractive interaction one can envisage that the trapped system undergoes a transition to the superfluid state. For instance $^6$Li atoms can be trapped in two different hyperfine states. In the spin polarized case the s-wave interaction turns out to be very strong and attractive (scattering length: $a$ = -2063 $a_0$ with $a_0$ the Bohr radius) favoring a phase transition to the superfluid state. This possibility has recently provoked a number of theoretical investigations (see [5] for a more detailed discussion of a possible superfluid state). One major question which is under debate is how to detect the superfluidity of such a fermionic system, since in contrast to a bosonic system the density of a Fermionic system is hardly affected by the transition to the superfluid state [6]. Several proposals such as the study of the decay rate of the gas or of the scattering of atoms off the gas have been advanced [5]. Though such investigations may give precious indications of a possible superfluid phase, we think that in analogy with nuclear physics, a measurement of the moment of inertia certainly would establish an unambiguous signature of superfluidity. To measure the spin and the rotational energy of trapped atoms definitely is a great challenge for the future. However, in nuclear physics, where $\gamma$-spectroscopy is extremely well developed, the strong reduction of the moment of inertia with respect to its rigid body value has been considered as a firm indicator of nucleon superfluidity immediately after the discovery of nuclear rotational states almost half a century back [6]. Therefore awaiting future experimental achievement also for trapped fermionic atoms, it is our intention in this work to give some theoretical estimates of the moment of inertia as a function of deformation of the traps or temperature of the gas. In this study we can largely profit from the experience nuclear physicists have accumulated over the last decades in describing such phenomena. The expectation is indeed that there will be a great analogy between the physics of confined atomic Fermions and what one calls in nuclear physics the liquid drop part of the nucleus. As astonishing as it may seem assemblies of fermions containing no more than ~200 particles (nucleons) already exhibit an underlying macroscopic structure well known from the Bethe-Weizsaecker formula for nuclear masses [6]. In superfluid rotating nuclei as early as 1959 Migdal proposed a statistical description of the nuclear moment of inertia [7] which grasped the essential physics of a self contained rotating superfluid Fermi liquid drop and which serves as a reference even today.

In the present work we will cast Migdal's approach into the more systematic language of the Thomas Fermi theory which together with its extensions is applied since long to normal fluid but also to superfluid nuclei [6,8,9]. It is fortunate that we can profit from this experience for the description of trapped fermions, since their number of order $10^5$, together with the smoothness of the potential, certainly turns a statistical description into a very precise tool. On the other hand it may not be excluded that in the future the study of much smaller systems of trapped atomic Fermions with numbers ~$10^2$ may be studied probably revealing many analogies with nuclei such as shell structure etc. The investigation of the transition from microscopic to macroscopic as the number of particles is increased continuously may then become a very interesting field also in the case of atomic Fermions. In detail our paper is organized as follows. In sections 2 and 3 we review the Thomas Fermi approach to inhomogeneous superfluid Fermi systems. In section 4 first the so-called Inglis part of the moment of inertia of a rotating superfluid and confined gas of atomic Fermions is presented. Second the influence of the reaction of the pair field on the moment of inertia is calculated. It is shown that this leads to the irrotational flow value in the limit of strong pairing. In section 5 the current distributions in the superfluid and normal fluid regimes are contrasted. In section 6 the numerical results are presented in detail together with discussions and conclusions.

## 2. Thomas Fermi approach to fermionic atoms in deformed traps

The Thomas-Fermi approach (TF) to trapped gases of atomic Bosons is a well accepted practice [2]. For trapped atomic Fermions the same approximation underlies different conditions. It has, however, recently also been applied to this kind of situation [5]. The TF approach for Fermions also is extensively applied to other finite systems such as atomic nuclei, metallic clusters, etc. The smallness parameter is given by

$$\eta = \frac{\nabla V(r)}{k_F(r) V(r)}. \tag{2.1}$$

Where V is the mean field potential and $k_F(r)$ the local Fermi momentum,

$$k_F(r) = \sqrt{\frac{2m}{\hbar^2}(\varepsilon_F - V(r))}. \tag{2.2}$$

With a typical frequency of the external harmonic potential of $\omega_0 = 7\,\text{nK}$ and $\varepsilon_F = 600$ nK one realizes that $\eta \ll 1$ up to close to the end of the classically allowed region. For integrated quantities the region around the classical turning point carries little weight and therefore the TF approximation for a number of trapped atoms of the order of $10^5$ is certainly very well justified.

Furthermore, as in the boson case the TF approach [6] to trapped atomic gases becomes extremely simple by the fact that the large interparticle distance makes a pseudopotential approximation to atomic interactions valid. Let us therefore write down the TF equation for a doubly spin polarized system of trapped ($^6$Li) atoms in the normal fluid state. For convenience we first consider the system at zero temperature T discussing the T ≠ 0 case later on. In TF approximation the distribution function for particles in each spin state is given by (in this work we only consider equal occupation of both spin states)

$$f(\mathbf{R},\mathbf{p}) = \theta(\mu - H_{cl}), \tag{2.3}$$

with

$$H_{cl} = \frac{p^2}{2m} + V_{ex}(\mathbf{R}) - g\rho(\mathbf{R}). \tag{2.4}$$

Here $\mu$ is the chemical potential, and $V_{ex}(\mathbf{R})$ stands for the trap potential which is supposed to be of harmonic form. The density $\rho(\mathbf{R})$ is obtained from the self-consistency equation,

$$\rho(\mathbf{R}) = \int \frac{d^3 p}{(2\pi\hbar)^3} f(\mathbf{R},\mathbf{p}) = \frac{1}{6\pi^2} k_F^3(\mathbf{R}), \qquad (2.5)$$

with,

$$k_F(\mathbf{R}) = \sqrt{\frac{2m}{\hbar^2}(\mu - V_{ex}(\mathbf{R}) + g\rho(\mathbf{R}))}, \qquad (2.6)$$

the local Fermi momentum. The coupling constant g is related to the scattering length in the same way as in the case of Bose condensed gases [1,2] via

$$g = \frac{4\pi\hbar^2 |a|}{m}. \qquad (2.7)$$

The TF equation (2.5) leads to a cubic equation for the self-consistent density, which straightforwardly can be solved as a function of the external potential. In this paper our main interest will be the study of the moment of inertia of a rotating condensate. Since the study is very much simplified assuming that the self consistent potential is again a harmonic oscillator and since the effect of the attractive interaction between the atoms essentially results in a narrowing of the self consistent potential with respect to the external one we will use instead of the exact TF solution for the density the following trial ansatz for the local Fermi momentum

$$k_F^{trial}(\mathbf{R}) = \sqrt{\frac{2m}{\hbar^2}\left(\mu - \frac{m}{2}\left(\omega_x^2 R_x^2 + \omega_y^2 R_y^2 + \omega_z^2 R_z^2\right)\right)}, \qquad (2.8)$$

where $\omega_x, \omega_y$ and $\omega_z$ are the variational parameters. The chemical potential is determined from the particle number condition

$$N = \int d^3 r \rho^{trial}(r), \qquad (2.9)$$

and the kinetic energy density is given by

$$\tau(\mathbf{R}) = \int \frac{d^3 p}{(2\pi\hbar)^3} \frac{p^2}{2m} f(\mathbf{R},\mathbf{p}) = \frac{1}{10\pi^2}\left[k_F^{trial}(\mathbf{R})\right]^5. \qquad (2.10)$$

We then can analytically calculate the total energy

$$E(\omega_x, \omega_y, \omega_z) = \int d^3 R \left[\tau(\mathbf{R}) + V_{ex}(\mathbf{R})\rho(\mathbf{R}) - \frac{g}{2}\rho^2(\mathbf{R})\right], \qquad (2.11)$$

as a function of $\omega_x, \omega_y$ and $\omega_z$. Minimizing this expression with respect to $\omega_x, \omega_y$ and $\omega_z$ for a given external deformed harmonic oscillator potential,

$$V_{ex} = \frac{m}{2}\left(\omega_{0x}^2 R_x^2 + \omega_{0y}^2 R_y^2 + \omega_{0z}^2 R_z^2\right), \qquad (2.12)$$

leads to the variational solution. For the spherical case $\omega_x = \omega_y = \omega_z = \omega$, this is shown in Fig. 1. We see that this approximation to the TF equation is quite reasonable. For an external harmonic potential with frequency $\nu = \omega_0/2\pi = 144\,\mathrm{Hz}$ or $\hbar\omega_0/k_B = 6.9\,\mathrm{nK}$, corresponding to the conditions of the experiment of Bradley et al. [10], the variational frequency is $\hbar\omega/k_B = 7.69\,\mathrm{nK}$. Since $\omega > \omega_0$ this implies a compression of the density. Increasing $\omega$ by 6% ($\hbar\omega/k_B = 8.21\,\mathrm{nK}$) from its variational value allows an almost perfect reproduction of the full TF solution. We will adopt this latter value in all our forthcoming calculations. The experimental situation for the rotating deformed case is such that the rotation of the trap is performed around the x-axis (the long axis), permitting slight triaxial deformations in the plane perpendicular to the x-axis that is in the y-z plane. In order to simulate such an experimental situation we simply first make a volume conserving

$$(\omega_\perp)^2 \omega_x = \omega^3, \tag{2.13}$$

prolate deformation around the x-axis,

$$\sigma = \frac{\omega_x}{\omega_\perp} \tag{2.14}$$

$$\omega_\perp = \omega\sigma^{-(1/3)} \;;\; \omega_\perp = \omega_y = \omega_z$$
$$\omega_x = \omega\sigma^{(2/3)} \tag{2.15}$$

In order to increase the central density there is interest to make rather strong eccentricities and $\omega_x/\omega_\perp = \frac{1}{8}$ is a typical value which we will adopt in this paper.

In a second step we fix $\omega_x$ and deform around the x-axis again keeping the volume fixed. We define the deformation parameter as

$$\delta = \frac{\omega_z}{\omega_y}. \tag{2.16}$$

We finally have the two-parameter deformation

$$\omega_x = \omega \sigma^{(2/3)}$$
$$\omega_y = \omega \sigma^{-(1/3)} \delta^{-(1/2)} \tag{2.17}$$
$$\omega_z = \omega \sigma^{-(1/3)} \delta^{(1/2)}$$

with $\sigma \ll \delta < 1$. From now on we therefore will use for the nonsuperfluid Wigner function at zero temperature the expression

$$f(\mathbf{R},\mathbf{p}) = \theta\left(\mu - \frac{p^2}{2m} - V(\mathbf{R})\right), \tag{2.18}$$

with

$$V(\mathbf{R}) = \frac{m}{2}\left(\omega_x^2 R_x^2 + \omega_y^2 R_y^2 + \omega_z^2 R_z^2\right), \tag{2.19}$$

and $\omega_x$, $\omega_y$, $\omega_z$ from (2.17) and $\mu$ determined from the particle number condition.

## 3. The superfluid case

Since, as described in the introduction, trapped spin polarized $^6$Li atoms, in different hyperfine states, feel a strong attractive s-wave interaction, the system very likely will undergo a transition to the superfluid state at some critical temperature $T_c$ as was discussed in detail in ref. [5]. As we have pointed out in the introduction the superfluid state will unambiguously reveal itself in its value of the moment of inertia. At the moment the measurement of angular momenta of trapped Bose or Fermi gases has not been achieved and represents a future challenge to the experimenters. In order to establish how the two essential system parameters which are the value of the gap, the temperature of the system, and the deformation of the external trap influence the value of the moment of inertia, we will now proceed to its evaluation in the superfluid state.

Since we are dealing with an inhomogeneous system, even in the nonrotating case the gap is actually a nonlocal quantity $\Delta(\mathbf{r},\mathbf{r}')$ or in Wigner space $\Delta(\mathbf{R},\mathbf{p})$. We will find later that at zero temperature the coherence length of the Cooper pair, $\xi = \frac{\hbar^2}{m}\frac{k_F}{\Delta}$, is larger than the oscillator length $\ell = \sqrt{\hbar/m\omega_0} \approx 0.63 \times 10^5 a_0$ with $a_0$ the Bohr radius. We therefore have to be careful with applying the TF theory for temperatures T much lower than the critical temperature $T_c$ where the gap vanishes. We will discuss this point more thoroughly later and in the appendix. We therefore go on and apply the TF approximation to the superfluid state. It has been shown in [9,6] that to lowest order in $\hbar$ the gap equation is given by

$$\Delta(\mathbf{R},\mathbf{p}) = \int \frac{d^3k}{(2\pi\hbar)^3} v(\mathbf{p}-\mathbf{k}) \frac{\Delta(\mathbf{R},\mathbf{k})}{2E(\mathbf{R},\mathbf{k})} \tanh\left(\frac{E(\mathbf{R},\mathbf{k})}{2T}\right), \tag{3.1}$$

where $E(\mathbf{R},\mathbf{p})$ is the quasiparticle energy,

$$E(\mathbf{R},\mathbf{p}) = \sqrt{\left[\frac{p^2 - p_F^2(\mathbf{R})}{2m^*(\mathbf{R})}\right]^2 + \Delta^2(\mathbf{R},\mathbf{p})}, \qquad (3.2)$$

with $p_F(\mathbf{R}) = \hbar k_F(\mathbf{R})$ the local Fermi momentum (2.6). Since the effective mass $m^*$ is so far unknown for trapped gases of atomic Fermions we will take $m^* = m$. Furthermore, for the time being, as in [5], we will eliminate the interatomic potential v in (3.1), expressing it by the scattering length (2.7). We then obtain [5]

$$\Delta(\mathbf{R},\mathbf{p}) = g \int \frac{d^3k}{(2\pi\hbar)^3} \left[\frac{\tanh\left(\frac{E(\mathbf{R},\mathbf{k})}{2T}\right)}{2E(\mathbf{R},\mathbf{k})} - \frac{P}{2(\varepsilon_k - \varepsilon_F(\mathbf{R}))}\right] \Delta(\mathbf{R},\mathbf{k}), \qquad (3.3)$$

where P stands for principal value, $\varepsilon_k = \frac{\hbar^2 k^2}{2m}$ and $\varepsilon_F = \frac{\hbar^2 k_F^2}{2m}$. At zero temperature, as described in [11], (3.3) can be solved analytically in the limit $\frac{\Delta(\mathbf{R}, p_F(\mathbf{R}))}{\varepsilon_F(\mathbf{R})} \to 0$. The result is given by

$$\Delta_F(\mathbf{R}) \equiv \Delta(\mathbf{R}, k_F(\mathbf{R})) = 8e^{-2} \varepsilon_F(\mathbf{R}) e^{-\frac{\pi}{2k_F(\mathbf{R})|a|}}. \qquad (3.4)$$

A posteriori one can verify that $\frac{\Delta_F}{\varepsilon_F} \ll 1$ for all values of R and therefore (3.4) is an excellent approximation to (3.3). This also has been found in [5]. For 286500

$^6$Li atoms, the case considered in [5], the gap is shown for a spherical trap as a function of the radius in Fig. 2.

For the determination of the critical temperature $T_c$ and, later on, for the moment of inertia we will need the value $\Delta$ of the gap at the Fermi energy. Since the detailed level structure at the Fermi energy is unknown and in fact unimportant, we will consider the gap $\Delta(\varepsilon_F)$ averaged over the states at the Fermi energy

$$\Delta(\varepsilon_F) \equiv \Delta = Tr(\hat{\Delta}\hat{\rho}(\varepsilon_F)), \tag{3.5}$$

with

$$\hat{\rho}(\varepsilon) = \frac{1}{g(\varepsilon)} \sum_n |n\rangle\langle n| \delta(\varepsilon - \varepsilon_n) = \frac{1}{g(\varepsilon)} \delta(\varepsilon - H), \tag{3.6}$$

where $|n\rangle$ and $\varepsilon_n$ are the states and energies of the harmonic oscillator with frequency $\omega$ and

$$g(\varepsilon) = \sum_n \delta(\varepsilon - \varepsilon_n) = Tr\delta(\varepsilon - H), \tag{3.7}$$

is the level density.

It has been shown in [12] that again the TF approximation leads to an excellent average value

$$\Delta = \frac{1}{g^{TF}(\varepsilon_F)} \int \frac{d^3R\, d^3p}{(2\pi\hbar)^3} \Delta_F(R) \delta(\varepsilon_F - H_{cl}), \tag{3.8}$$

In the spherical case with $\Delta_F(R)$ from (3.4) all integrals but the radial one can be performed analytically, the latter being done numerically. For the case shown in Fig. 2 one obtains

$$\Delta = 16.4 \text{ nK}. \tag{3.9}$$

Quantum mechanically the BCS equations should be solved in the self-consistent Hartree-Fock (HF) basis and then $T_c$ is a global parameter which must be determined from the solution of the quantum mechanical gap equation. Since we believe that the value in (3.9) comes rather close to the quantum mechanical value of the gap at the Fermi energy, we can obtain $T_c$ from the usual BCS weak coupling relation [13] $\Delta = 1.76 T_c$ to be

$$T_c \approx 10 \text{ nK}. \tag{3.10}$$

From (3.9) we obtain the coherence length $\xi = \dfrac{\hbar^2}{m}\dfrac{k_F}{\Delta} = \dfrac{\varepsilon_F}{\Delta}\dfrac{2}{k_F}$. With $\varepsilon_F = 983.67$ nK which corresponds to our approximate "self-consistent" harmonic solution with $\omega = 8.21 \text{ nK}$ and $k_F|a| = 0.56$ one obtains $\xi \approx 4.10^5 a_0$ which is about a factor of seven larger than the oscillator length of the trap (see above) which contains 286500 particles. This seems to invalidate the TF-approximation. We, however, know by experience that often the TF-approximation remains quite reasonable beyond its limit [6]. For example the conditions of validity in [9,12] for superfluid nuclei are much worse than here and still the results are accurate beyond expectation. We therefore think that the values (3.9, 3.10) are reasonable estimates for the gap and the critical temperature. In order to check this assumption we give in the appendix a more refined semi-classical solution of the gap equation which only demands that the TF-approximation in the normal fluid state is well justified. We find values for $\Delta$ and $T_c$ which are ~ 30% lower than in (3.9)(3.10). In view of the crudeness of the TF-approach this indicates a quite satisfying consistency between the results.

We also will have to know the detailed T-dependence of the gap $\Delta(T)$ which however, in BCS theory, given $\Delta(0)$ and $T_c$, is determined by the universal function $\frac{\Delta(T)}{\Delta(0)}$ in terms of $\frac{T}{T_c}$. This function is determined from the solution of the equation [13]

$$-\ln\left(\frac{\Delta(T)}{\Delta(0)}\right) = A\left(\frac{\Delta(T)}{T}\right),$$

with (3.11)

$$A(u) = \int_0^\infty dy \frac{1}{\sqrt{y^2+u^2}}\left(1-\tanh\left(\frac{\sqrt{y^2+u^2}}{2}\right)\right).$$

For completeness it is shown in fig. 3. This T-dependence of the gap we will later use for the evaluation of the moment of inertia.

## 4. Moment of inertia

The moment of inertia of a rotating nucleus has fully been formulated in linear response theory (i.e. RPA) by Thouless and Valatin [14]. The corresponding expression is therefore called, in the nuclear physics literature [6], the Thouless-Valatin moment of inertia. It consists of two parts, the so-called Inglis term, which describes the free gas response, and the part, which accounts for the reaction of the mean field and pair potential to the rotation. In the superfluid case the Inglis part has been generalized by Belyaev [15] and the linear reaction of the gap parameter onto the value of the moment of inertia was first evaluated, together with the Inglis term, by Migdal [7]. The reaction of the HF field on the rotation is a minor effect and we will neglect it in this work. We therefore will write the moment of inertia as a sum of the Inglis-Belyaev term $\Theta_{I-B}$ and the Migdal term $\Theta_M$. In total

$$\Theta = \Theta_{I-B} + \Theta_M . \qquad (4.1)$$

In order to derive an expression for $\Theta$ in linear response theory we will use the Gorkov approach described in detail in many text books (in what follows we will use the notation of [16]). Since in addition the derivation of the linear response for $\Theta$ is given in the original article of Migdal [7] and rerepresented in a more elaborate version in [8], we will be very short here and only give more details where in our opinion the presentations in [7,8] may not be entirely explicit. Let us start writing down the Gorkov equations in matrix notation

$$\left(-\frac{\partial}{\partial \tau} - H + \mu \right)G = 1 - \Delta F^+$$

$$\left(\frac{\partial}{\partial \tau} + H^* + \mu \right)F^+ = \Delta^* G$$

(4.2)

with

$$H = H_0 - \Omega L_x \equiv H_0 + H_1,$$

(4.3)

where now $H_0$ is the shell model Hamiltonian (2.4) or rather the approximate one used in (2.18), (2.19) and

$$L_x = r_y p_z - r_z p_y,$$

the angular momentum operator corresponding to a rotation with angular frequency $\Omega$ around the x-axis. In (4.2) G and F are the normal and anomal Matsubara Green's functions (see Chap. 51 of [15])

$$G_{nn'} = -\left\langle T_\tau a_n(\tau) a_{n'}^+(\tau') \right\rangle$$

$$F^+_{nn'} = -\left\langle T_\tau a_n^+(\tau) a_{n'}^+(\tau') \right\rangle$$

(4.4)

Linearising (4.2) with respect to $H_1$, that is $G = G_0 + G_1$, $F^+ = F_0^+ + F_1^+$ and $\Delta = \Delta_0 + \Delta_1$ (as mentioned we will neglect the influence of the rotational field on $H_0$) one obtains for (4.2)

$$G_1 = G_{1I-B} + G_{1M}, \tag{4.5}$$

with

$$G_{1I-B} = G_0 H_1 G_0 + F_0^+ H_1^* F_0^+$$

$$\tag{4.6}$$

$$G_{1M} = -G_0 \Delta_1 F_0^+ - F_0^+ \Delta_1^* G_0$$

and

$$F_1^+ = D_0 H_1^* F_0^+ + F_0^+ H_1 G_0 - F_0^+ \Delta_1 F_0^+ + D_0 \Delta_1^* G_0, \tag{4.7}$$

where,

$$D = \frac{i\omega_n - H_0}{\omega_n^2 + H_0^2 + \Delta_0^2}, \quad G_0 = \frac{i\omega_n + H_0}{\omega_n^2 + H_0^2 + \Delta_0^2}, \quad F_0^+ = \frac{\Delta_0}{\omega_n^2 + H_0^2 + \Delta_0^2},$$

and $\omega_n$ are the Matsubara frequencies [16].

In (4.5,4.7) we have split the first order Green's function in an obvious notation into the Inglis-Belyaev and Migdal contributions. For the latter one needs the linear reaction of the pair field to the rotation. We later will see how this can be determined from (4.7). First let us, however, evaluate the I-B part of the moment of inertia.

## 4.1 The Inglis-Belyaev part of the moment of inertia

The I-B part of the moment of inertia can be evaluated without the knowledge of $\Delta_1$ i.e. without the use of (4.7). The density response corresponding to $G_{1I-B}$ of (4.5) is evaluated from the limit $\tau' \to \tau^+$ or from summing over the Matsubara frequencies in the upper half plane (see Ch. 7 of [16]) . One obtains the well known result [6, 7, 8, 15,16]

$$(\rho_{1I-B})_{nn'} = \langle n|L_x|n'\rangle F_{nn'}, \tag{4.8}$$

with

$$F = F_+(1 - f - f') + F_-(f - f'), \tag{4.9}$$

where

$$F_\pm = F_\pm(\varepsilon_n, \varepsilon_{n'}) = \frac{E_n E_{n'} \mp \xi_n \xi_{n'} - \Delta(\varepsilon_n)\Delta(\varepsilon_{n'})}{2 E_n E_{n'}(E_n \pm E_{n'})}, \tag{4.10}$$

$$f = f(\varepsilon_n) = \frac{1}{1 + e^{E_n/T}} \quad ; \quad f' = f(\varepsilon_{n'}) \tag{4.11}$$

and

$$E_n = \sqrt{\xi_n^2 + \Delta^2(\varepsilon_n)} \quad ; \quad \xi_n = \varepsilon_n - \mu, \tag{4.12}$$

are the quasiparticle energies with as ingredient $\varepsilon_n$, the energies of the harmonic oscillator potential (2.19). The gap parameters $\Delta_n$ have been replaced in (4.10), in analogy to (3.5), to statistical accuracy by $\Delta(\varepsilon_n)$, the ones averaged over the energy shell. The moment of inertia is given by

$$\Theta_{I-B} = Tr(L_x \rho_{1I-B}). \tag{4.13}$$

Since we are interested at temperatures $T \leq T_c$, which are very low with respect to the Fermi energy, we checked that one can to very good accuracy neglect in (4.9) the thermal factors (4.11). The only important temperature dependence of the moment of inertia therefore exists via the T-dependence of the gap. We thus will henceforth treat all formulas as in the T=0 limit keeping, however, the T-dependence of the gap. With this in mind we can write for the moment of inertia

$$\Theta_{I-B} = \sum_{nn'} \iint d\omega d\omega' \delta(\omega - \varepsilon_n) \delta(\omega' - \varepsilon_{n'}) |\langle n|L_x|n'\rangle|^2 F_+(\omega, \omega'), \tag{4.14}$$

In this formula the important quantity to calculate to statistical accuracy is,

$$L_x^2(n,n') \equiv |\langle n|L_x|n'\rangle|^2 = Tr[(L_x)(\ |n'\rangle\langle n'|L_x|n\rangle\langle n|\ )]$$
$$= \int \frac{d^3R d^3p}{(2\pi\hbar)^3}(L_x)_W\ [|n'\rangle\langle n'|L_x|n\rangle\langle n|]_W \quad (4.15)$$

where $O_W \equiv O(\mathbf{R},\mathbf{p})$ means the Wigner transform of the operator $O$ [6]. To this purpose we again replace the density matrices $|n\rangle\langle n|$ and $|n'\rangle\langle n'|$ by their average on the energy shell (3.6)

$$|n\rangle\langle n| \to \hat{\rho}(\varepsilon_n).$$

We therefore obtain

$$\Theta_{I-B} = \iint d\omega d\omega' \int \frac{d^3R d^3p}{(2\pi\hbar)^3}[(L_x)_W (L_x(\omega,\omega'))_W\ ]F_+(\omega,\omega'), \quad (4.16)$$

with

$$(L_x(\omega,\omega'))_W = [\delta(\omega'-\hat{H}_0)\hat{L}_x\delta(\omega-\hat{H}_0)]_W. \quad (4.17)$$

Introducing into (4.17) the Fourier representations of the two $\delta$-functions and transforming to center of mass and relative coordinates one obtains

$$(L_x(\omega,\omega'))_W = \iint \frac{dTd\tau}{(2\pi\hbar)^2} e^{2iET} e^{i\varepsilon\tau/2}\left[e^{-iH_0T}L_x\left(\frac{\tau}{2}\right)e^{-iH_0T}\right]_W, \quad (4.18)$$

with

$$E = \frac{\omega+\omega'}{2}\ ;\quad \varepsilon = \omega-\omega', \quad (4.19)$$

and

$$O(t) = e^{iH_0 t} O(0) e^{-iH_0 t}. \tag{4.20}$$

To lowest order in $\hbar$ we replace the triple operator product in (4.18) by the product of their Wigner transforms [6]

$$\lim_{\hbar \to 0} \left[ e^{-iH_0 T} L_x\left(\frac{\tau}{2}\right) e^{-iH_0 T} \right]_W = e^{-i2H_{0cl}T} L_x^{cl}\left(\frac{\tau}{2}\right), \tag{4.21}$$

and therefore

$$(L_x(\omega, \omega'))_W = L_x(E, \varepsilon, \mathbf{R}, \mathbf{p}) = \frac{1}{2} \delta(E - H_{0cl}) \int \frac{d\tau}{(2\pi\hbar)} e^{\frac{i}{\hbar}\varepsilon\frac{\tau}{2}} L_x^{cl}\left(\frac{\tau}{2}\right), \tag{4.22}$$

with

$$L_x^{cl}(t) = R_y(t) p_z(t) - R_z(t) p_y(t). \tag{4.23}$$

At this point the choice of our approximate self consistent potential of harmonic oscillator form (see 2.19) turns out to be very helpful, since the classical trajectories in (4.22) can be given analytically

$$R_i(t) = R_i \cos(\hbar \omega_i t) + \frac{p_i}{m \omega_i} \cos(\hbar \omega_i t)$$

$$p_i(t) = p_i \cos(\hbar \omega_i t) + m \omega_i R_i \cos(\hbar \omega_i t) \tag{4.24}$$

with $i = x, y, z$.

In the phase space integral of (4.15), for reasons of symmetry, only the diagonal terms of $L_x^{cl} . L_x^{cl}(t)$ survive and therefore we obtain

$$\int \frac{d^3R\, d^3p}{(2\pi\hbar)^3} L_x^{cl} L_x^{cl}(t) = \int d^3R\, \rho^{TF}(\mathbf{R})\left(R_y^2 + R_z^2\right)\cos(\hbar\omega_y t)\cos(\hbar\omega_z t)$$
$$+ \left(R_y^2 \frac{\omega_y}{\omega_z} + R_z^2 \frac{\omega_z}{\omega_y}\right)\sin(\hbar\omega_y t)\sin(\hbar\omega_z t) \tag{4.25}$$

where

$$\rho^{TF} = \frac{1}{6\pi^2}\left[\frac{2m}{\hbar^2}(E-V)\right]^{3/2}, \tag{4.26}$$

is the density in TF approximation (see eq. (2.5)).

The product of cosine and sine in (4.25) can be expressed in terms of the cosine of the sum and difference of the arguments and then the $\tau$-integral in (4.23) can be performed. This leads to $\delta$-functions which allows to perform also the $\varepsilon$-integral. Furthermore, as shown by Migdal [7]

$$F_+(E,\varepsilon) \approx \left[1 - G\left(\frac{\varepsilon}{2\Delta}\right)\right]\delta(E-\mu), \tag{4.27}$$

where (see eq. (3.5)),

$$\Delta = \Delta(\varepsilon_F),$$

and

$$G(x) = \frac{\operatorname{arsinh}(x)}{x\sqrt{1+x^2}}. \tag{4.28}$$

Finally one obtains for the I-B part of the moment of inertia the following analytical expression [7,8]

$$\Theta_{I-B} = \Theta_{rigid}\left[1 - \frac{G_+\omega_-^2 + G_-\omega_+^2}{\omega_-^2 + \omega_+^2}\right], \tag{4.29}$$

where

$$\omega_\pm = \omega_y \pm \omega_z \quad ; \quad G_\pm = G\left(\frac{\hbar\omega_\pm}{2\Delta}\right), \tag{4.30}$$

and

$$\Theta_{rigid} = \left(\frac{\mu^4}{24\hbar^3}\right)\frac{(\omega_y^2 + \omega_z^2)}{\omega_z^3 \omega_x \omega_y^3}, \tag{4.31}$$

is the moment of inertia of rigid rotation. From (4.29) we see that

$$\lim_{\Delta \to 0}\Theta_{I-B} = \Theta_{rigid} \quad ; \quad \lim_{\Delta \to \infty}\Theta_{I-B} = 0.$$

The latter result is clearly unphysical and we will see how the account of the reaction of the pair field on the rotation will reestablish the physical situation.

**4.2 The Migdal term**

The density response corresponding to the Migdal term is obtained from (4.6)

$$(\rho_{1M})_{n,n'} = \frac{\xi_n \Delta_{1nn'}\Delta_{0n'} + \Delta_{0n}\Delta^*_{1nn'}\xi_{n'}}{2E_n E_{n'}(E_n + E_{n'})}. \tag{4.32}$$

In (4.32) we need to know $\Delta_1$ which we can gain from (4.7) in the following way; in the limit $\tau' \to \tau^+$ we obtain from $F_1^+$ the anomal density $\kappa_1^+$,

$$\left(\kappa_1^+\right)_{nn'} = -\frac{\xi_n H_{1nn'}^* \Delta_{0n'} + \Delta_{0n} H_{1nn'} \xi_{n'} + \Delta_{0n} \Delta_{1nn'} \Delta_{0n'} - (E_n E_{n'} + \xi_n \xi_{n'}) \Delta_{1nn'}^*}{2 E_n E_{n'} (E_n + E_{n'})}. \qquad (4.33)$$

In analogy with the non-rotating case where $\kappa_0 = \frac{\Delta}{2E}$, we also have

$$\left(\kappa_1^+\right)_{nn'} = -\Delta_{1nn'}^* \left(\frac{1}{4E_n} + \frac{1}{4E'_{n'}}\right). \qquad (4.34)$$

This relation stems from the fact that the quasiparticle energies contain the gap only in the form $\Delta \Delta^*$ and therefore there is no further first order correction, since in our case the external field is a time odd operator and thus

$$\Delta_1^* = -\Delta_1 \equiv -i\Omega\chi. \qquad (4.35)$$

Equating (4.33) and (4.34) yields

$$\frac{2\xi_n H_{1nn'}^* \Delta_{0n'} + 2\Delta_{0n} H_{1nn'} \xi_{n'} + 2\Delta_{0n} \Delta_{1nn'} \Delta_{0n'} + \left(\Delta_{0n}^2 + \Delta_{0n'}^2 + (\xi_n - \xi_{n'})^2\right) \Delta_{1nn'}^*}{2 E_n E_{n'} (E_n + E_{n'})} = 0. \qquad (4.36)$$

At this point we again exploit the fact that expression (4.36) is strongly peaked around the Fermi energy surface. Following [7], in analogy with (4.27), we have

$$[E_n E_{n'} (E_n + E_{n'})]^{-1} \approx \frac{1}{\Delta^2} G\left(\frac{\varepsilon_n - \varepsilon_{n'}}{2\Delta}\right) \delta\left(\frac{\varepsilon_n + \varepsilon_{n'}}{2} - \mu\right). \qquad (4.37)$$

With (4.35) we then obtain for (4.36)

$$\left[\frac{\langle n|\dot{L}_x|n'\rangle}{2\Delta} + \left(\frac{\varepsilon_n - \varepsilon_{n'}}{2\Delta}\right)^2 \chi_{nn'}\right] G\left(\frac{\varepsilon_n - \varepsilon_{n'}}{2\Delta}\right) \delta\left(\frac{\varepsilon_n + \varepsilon_{n'}}{2} - \mu\right) = 0, \quad (4.38)$$

where $\dot{L}_x$ stands for the time derivative of $L_x$. Summing on n and n' and following exactly the same line of semi-classical approximations as the ones used for the derivation of $\Theta_{I-B}$ one arrives at the following relation [8]

$$\int_{-\infty}^{+\infty} d\tau G(\tau) \int \frac{d^3p}{(2\pi\hbar)^3} \left[\frac{\dot{L}_x^{cl}}{2\Delta} - \frac{\ddot{\chi}(\tau)}{4\Delta^2}\right] \delta(\mu - H_{0cl}) = 0 \quad (4.39)$$

where $G(\tau)$ is the Fourier transform of $G(x)$ (4.28).

For the potential in (2.19) (4.39) is solved by

$$\chi(\mathbf{R}) = \alpha R_y R_z, \quad (4.40)$$

with

$$\alpha = -2\Delta m \omega_+ \omega_- \frac{G_+ + G_-}{\omega_+^2 G_+ + \omega_-^2 G_-}. \quad (4.41)$$

Inserting this solution into (4.32) leads for the Migdal part of the moment of inertia to [7,8]

$$\Theta_M = \Theta_{rigid} \frac{\omega_+^2 \omega_-^2}{\omega_+^2 + \omega_-^2} \frac{(G_+ + G_-)^2}{\omega_+^2 G_+ + \omega_-^2 G_-}. \quad (4.42)$$

Together with (4.29) the expression for the moment of inertia is now complete. Let us again mention that we neglected the temperature dependence besides the one contained in $\Delta = \Delta(T)$, since all other T-dependence for $T < T_c$ is negligible. The moment of inertia can then be calculated as a function of deformation and temperature. For example it is immediately verified that for $\Delta \to \infty$ (4.42) yields the irrotational flow value

$$\lim_{\Delta \to \infty} \Theta_M = \Theta_{irrot} = \Theta_{rigid} \left( \frac{\omega_y^2 - \omega_z^2}{\omega_y^2 + \omega_z^2} \right)^2, \tag{4.43}$$

and therefore

$$\lim_{\Delta \to \infty} \Theta = \lim_{\Delta \to \infty} \left( \Theta_{I-Bt} + \Theta_M \right) = \Theta_{irrot}, \tag{4.44}$$

which is the correct physical result.

5. **Current distribution**

Other quantities, which may be interesting also from the experimental point of view, are the current distributions of the superfluid rotating gas. Indeed after a sudden switch off of the (rotating) trap the atoms will expand keeping memory of their rotational state. So if the velocity distribution of the expanding atoms can be measured, one may be able to deduce the rotational motion the atoms have had before the trap was taken away. The current distribution, as we will see, depends, as the moment of inertia, strongly on the superfluid state of the gas. In order to calculate the current distribution we first write down the Wigner function of the density response which can easily be read off from the formulas given in Section 4. In obvious notations we obtain [8]

$$\rho_{1I-B}(\mathbf{R},\mathbf{p}) = \Omega[\mathbf{R}\times\mathbf{p}]_x \delta(\mu - H_{0cl})$$

(5.1)

$$-\Omega\left[\frac{\omega_+ G_- - \omega_- G_+}{2\omega_z} R_y p_z - \frac{\omega_+ G_- + \omega_- G_+}{2\omega_y} R_z p_y\right]\delta(\mu - H_{0cl})$$

$$\rho_{1M}(\mathbf{R},\mathbf{p}) = \frac{\hbar^2}{2m}\frac{\alpha}{\Delta}\Omega\left[\frac{\omega_+ G_+ - \omega_- G_-}{\omega_z} R_y p_z + \frac{\omega_+ G_+ + \omega_- G_-}{\omega_y} R_z p_y\right]\delta(\mu - H_{0cl}),$$

(5.2)

With the usual definition of the current

$$\mathbf{j}(\mathbf{R}) = \int \frac{d^3 p}{(2\pi\hbar)^3}\frac{\mathbf{p}}{m}\rho(\mathbf{R},\mathbf{p}),$$

(5.3)

one obtains

$$j_y^{I-B} = -\rho_{TF}(\mathbf{R}) R_z \Omega\left[1 - \frac{\omega_+ G_- + \omega_- G_+}{2\omega_y}\right],$$

(5.4a)

$$j_z^{I-B} = -\rho_{TF}(\mathbf{R}) R_y \Omega\left[1 - \frac{\omega_+ G_- - \omega_- G_+}{2\omega_z}\right],$$

(5.4b)

$$j_y^M = -\rho_{TF}(\mathbf{R}) R_z \Omega\left[\frac{\omega_+\omega_-(G_- + G_+)}{\omega_+^2 G_+ + \omega_-^2 G_-}\frac{\omega_+ G_- + \omega_- G_+}{\omega_y}\right],$$

(5.5a)

$$j_z^M = -\rho_{TF}(\mathbf{R}) R_y \Omega\left[\frac{\omega_+\omega_-(G_- + G_+)}{\omega_+^2 G_+ + \omega_-^2 G_-}\frac{\omega_+ G_- - \omega_- G_+}{\omega_z}\right],$$

(5.5b)

with of course, $j_x = 0$.

Again we see that in the limit $\Delta \to \infty$ the current approaches the correct irrotational flow limit

$$\mathbf{j} \underset{\Delta \to \infty}{\longrightarrow} -2\rho_{TF}\Omega \frac{\omega_y^2 - \omega_z^2}{\omega_y^2 + \omega_z^2} \nabla(r_y r_z), \tag{5.6}$$

whereas in the limit of $\Delta \to 0$ we obtain a rigid body current. As we have seen for $\Theta$, as a function of temperature and deformation, we easily can go from one limit to the other.

6. **Results and conclusion**

We show in Fig. 4 a and b the current distribution for the two extreme cases of irrotational and rigid body flow in the laboratory frame respectively. We see that the flow pattern is completely different in the two cases. In Fig. 4b the flow pattern clearly corresponds to rigid rotation of an ellipsoid with the long axis in z-direction. Also Fig. 4a represents a typical irrotational flow pattern well know from hydrodynamics. As a function of temperature one continuously can pass from one flow pattern to the other. The point we want to make is that for small deformations $\delta$, as can be seen from (5.6) there is almost no irrotational current for low temperatures and this will then be reflected in a very low rotational energy as we will discuss now.

In Fig. 5 we show $\Theta$ as a function of $\Delta(T)$ and with Fig. 3 also as a function of T. We see that for a typical eccentricity $\delta = \frac{\omega_z}{\omega_y} = 0.8$ the moment of inertia changes, as a function of temperature, by large factors. At $T \approx 0$ the gap values found in this paper are in the range of 10-20 nK and therefore the moment of inertia is close to its irrotational flow limit. This actually means that the moment of inertia is, with respect to its rigid body value, very small, since for $\delta \to 1$, i.e. for spherical symmetry around the rotational axis (x-axis) the moment of inertia goes to zero (see (4.43, 4.44)). Consequently in this case the gas is not at all following the rotation of the trap. However, increasing the temperature i.e. decreasing the gap value has a dramatic influence on the rotational motion of the gas, since in the range $0 < T < T_c$ the moment of inertia raises very steeply attaining its rigid body value for $T = T_c$. In this limit the gas rotates as a whole with the same angular frequency as the trap. The abruptness of the raise is the more pronounced, the smaller the eccentricity $\delta$ is. (see Fig. 5). Experimentally non-destructive or expansion imaging can be used to watch the gas rotate and then the rotational energy

$$E_{rot} = \frac{\Theta}{2}\Omega^2,\tag{6.1}$$

can be obtained by integrating the angular velocity over the density profile[*]. The rotational energy therefore directly follows the variation of the moment of inertia. One deduces that the measurement of the variation as a function of T of the rotational energy should be well within the experimental possibilities, once the technique of putting the trap into rotation has been mastered.

In our discussion we have ignored the possibilities of vortex formation. The determination of the onset of instabilities versus vortex formation in a finite Fermi system is not a completely easy task and we will postpone such an investigation to future work.

---

[*] We are grateful to the referee for pointing to this possibility.

However, since the rotational frequencies $\Omega$ considered in this paper are much smaller than the oscillator constant $\omega_0$ ($\Omega/\omega_0 \ll 1$) we think that our result will not be spoiled by the appearance of vortices. An indication can also come from the case of trapped Bosons where vortices, depending somewhat on the number of atoms, do not appear for values $\Omega/\omega_0 < .5$ (see ref. [2])

From the above discussion we see that it may well be in experimental reach to reveal an eventual superfluid state of the gas once the technique of putting the trap into rotation will be mastered experimentally. A closely related phenomenon to rotation is the so-called scissors mode which was originally discussed and found in deformed nuclei [17] and then proposed [18] and also very recently found [19] for trapped Boson condensates. Suppose the trapped atomic system is rotating very slowly and suddenly the rotation of the deformed trap potential will be stopped. Due to inertia the atomic cloud will continue rotating back and forth around the fixed trap position if the initial rotation was gentle enough. If for the purpose of a rough argument we suppose that this oscillatory motion has so small amplitude that in a first approximation we can neglect shape distortions of the cloud, then, if the oscillations are in harmonic regime, the frequency of the scissors mode is given by

$$\omega_S = \sqrt{\frac{C}{\Theta}}, \qquad (6.2)$$

where C is the constant of the restoring force. The frequency $\omega_S$ will strongly depend on whether the system is in the superfluid state or not. In this way the above cited experiment has indeed unambiguously revealed that the Bose condensate is in a superfluid state [19]. It is evident that scissors modes could also be excited in trapped Fermi systems, as this was already mentioned in [18]. Since in Fermi systems for temperatures T~$T_c$ one can suppose that the temperature dependence of the force constant is weak with respect to the one of the moment of inertia $\Theta$, one will find a strong difference for the value of $\omega_S$ in the superfluid and unpaired regimes respectively. As long as the temperature is so small that the normal fluid component can be neglected, the temperature dependence of $\omega_S$ can be deduced from the one of $\Theta$ obtainable from Figs 3 and 5 of this work. We are, of course, aware that the experimental situation may be more complicated needing a more refined discussion similar to the one given in [18]. A more detailed investigation of the scissors mode for trapped Fermions may be given in future work.

In summary we proposed in this work to measure the dynamics of a rotating trapped gas of atomic Fermions as a function of temperature and deformation to detect whether the system is in a superfluid state or not. Quite detailed and quantitative calculations for the moment of inertia and velocity distributions have been presented. Other quantities well studied in the case of rotating superfluid nuclei [6] such as Yrast lines, even-odd effects, particle alignment, etc., may also become of interest in this case.

Acknowledgements: We gratefully acknowledge very useful discussions with C. Gignoux , D. Guéry-Odelin, S. Stringari and W. Zwerger and a careful reading of the manuscript by M. Durand. One of us (X.V) also acknowledges financial support from DGCYT (Spain) under grant PB95-1249 and from the DGR (Catalonia) under grant GR94-1022.

## Appendix

In this appendix we want to give a more refined semi-classical solution of the gap equation. Let us write the quantal version of (3.3) at T = 0 in BCS approximation [6]

$$\Delta_n = \sum_{n'} \langle n\bar{n}|v|n'\bar{n}'\rangle \Delta_{n'} \left[ \frac{\tanh\left(\frac{E_{n'}}{2T}\right)}{2E_{n'}} - \frac{P}{2(\varepsilon_{n'} - \varepsilon_F)} \right], \quad (A.1)$$

where n labels the states of the (spherical) harmonic oscillator with single particle energies $\varepsilon_n$ and $\bar{n}$ is the time reversed state. As usual $E_n = \sqrt{(\varepsilon_n - \varepsilon_F)^2 + \Delta_n^2}$ is the quasiparticle energy and $\langle n\bar{n}|v|n'\bar{n}'\rangle$ is the matrix element of the zero range two body force

$$g\delta(\mathbf{r} - \mathbf{r}'). \quad (A.2)$$

Since what matters is the gap at the Fermi level and since for particle numbers of the order $10^5$ the degeneracy of the oscillator shells is very high it seems a very reasonable approximation to replace in (A.1) all quantities by their corresponding values averaged over the energy shell (3.5, 3.6). Equation (A.1) can then be written as

$$\Delta_n = \int_0^\infty d\varepsilon' g(\varepsilon') v(\varepsilon, \varepsilon') \Delta(\varepsilon') \left[ \frac{\tanh\left(\frac{E(\varepsilon')}{2T}\right)}{2E(\varepsilon')} - \frac{P}{2(\varepsilon' - \varepsilon_F)} \right], \quad (A.3)$$

where $g(\varepsilon)$ is the level density (3.7) and $v(\varepsilon,\varepsilon')$ is the averaged two body matrix element

$$v(\varepsilon,\varepsilon') = \frac{1}{g(\varepsilon)g(\varepsilon')} \sum_{n,n'} \delta(\varepsilon - \varepsilon_n) \delta(\varepsilon' - \varepsilon_{n'}) \langle n\bar{n}|v|n'\bar{n}'\rangle . \quad \text{(A.4)}$$

At this stage one could try to solve the gap equation numerically. However, again in view of the huge number of particles it is certainly a good approximation to pass to the Thomas-Fermi limit. For the level density $g(\varepsilon)$ this is immediate. The TF limit of (A.4) can be obtained in locally summing over plane waves and we obtain

$$v(\varepsilon,\varepsilon') = \frac{g}{g^{TF}(\varepsilon)g^{TF}(\varepsilon')} \left(\frac{2m}{\hbar^2}\right)^3 \frac{1}{4\pi^3} \int_0^{\inf(r_\varepsilon, r_{\varepsilon'})} dr\, r^2 \sqrt{\varepsilon - V(r)}\sqrt{\varepsilon' - V(r)}, \quad \text{(A.5)}$$

where $r_\varepsilon$ is the classical turning point given by $\varepsilon = V(r_\varepsilon)$ and $V(r) = m\omega^2 r^2/2$ is the harmonic oscillator potential. We have made numerical check that (A.5) is indeed a good approximation to the quantal counterpart for the case of large particle numbers [20]. We notice that (A.5) only needs the TF approximation in the nonsuperfluid state where it is well justified (see sect. 2). Having an expression for average level density and matrix element at hand we can proceed to solve (A.3) We will do this again in the limit $\Delta(\varepsilon_F)/\varepsilon_F \ll 1$ and obtain (see [11]) at T = 0

$$\Delta(\varepsilon_F) = 8\varepsilon_F e^{\left\{-\frac{15\pi^2}{64k_F|a|} + I(\varepsilon_F)\right\}}, \quad \text{(A.6)}$$

with

$$I(\varepsilon_F) = 2\int_0^1 dx \frac{x^5 \frac{v^2(x^2\varepsilon_F, \varepsilon_F)}{v^2(\varepsilon_F, \varepsilon_F)} - 1}{1 - x^2}.$$ (A.7)

The integral $I(\varepsilon_F)$ is evaluated numerically and we obtain

$$\Delta(\varepsilon_F) = 8\varepsilon_F e^{-2.447} e^{-\frac{15\pi^2}{64 k_F |a|}}.$$ (A.8)

With $\varepsilon_F = 983.67$ nK which corresponds to our "self-consistent" harmonic solution and $k_F |a| = .56$ one obtains $\Delta(\varepsilon_F) = 11.29$ nK. This value is about 30% smaller than the one extracted in (3.9), which, however, in view of the roughness of the TF approximation can be considered as a rather satisfying consistency of the results.

# Figure captions

**Figure 1 :** Density profiles, for the case of a spherical trap, of the non-interacting case (full line), the interacting case once calculated exactly from (2.5) (crosses) with $V_{ex}$ given by (2.12) and once using the variationally determined harmonic oscillator potential (open squares). Squeezing the variational $\omega$ by 6% yields a density which lies on top of the exact TF solution.

**Figure 2 :** The gap for a spherical trap as a function of the radius

**Figure 3 :** Ratio of the energy gap to the gap at T=0°K as a function of temperature

**Figure 4 a and b :** The current distribution for the two extreme cases of irrotational (a) and rigid body (b) flow in the laboratory frame. In both cases the deformation parameters are set to ($\sigma = \frac{1}{8}, \delta = .8$) and the angular frequency $\Omega$ around the x-axis to 1 nK.

**Figure 5 :** The moment of inertia as a function of the gap for different values of the deformation $\delta = \frac{\omega_z}{\omega_y}$ and $\sigma = \frac{1}{8}$.

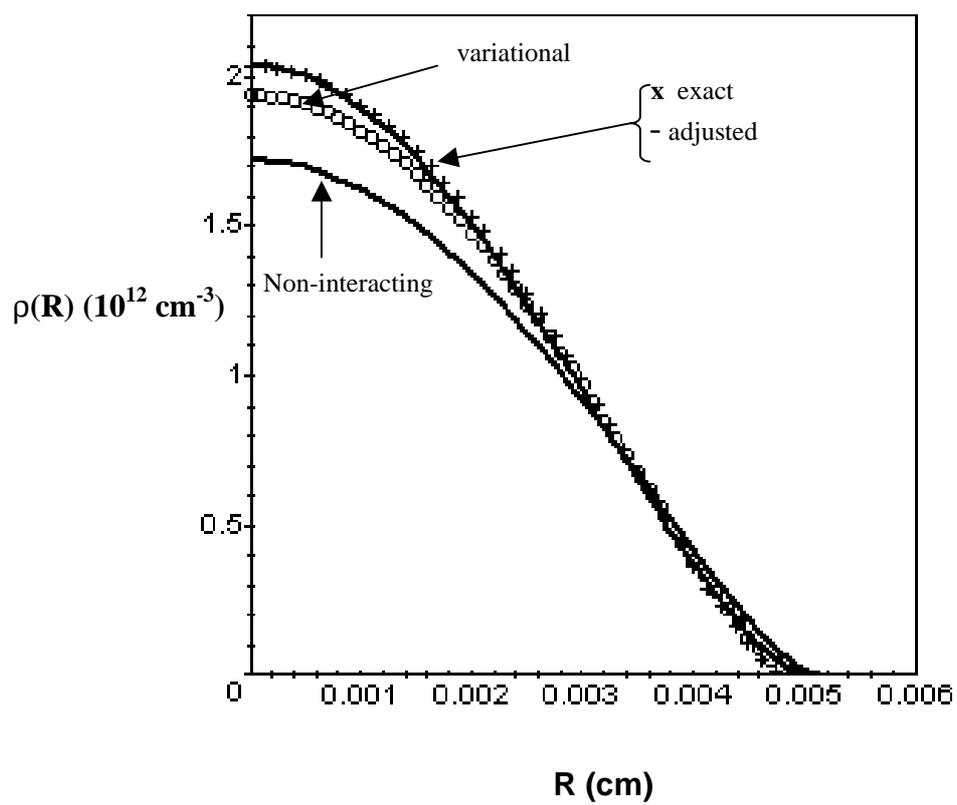

**Figure 1**

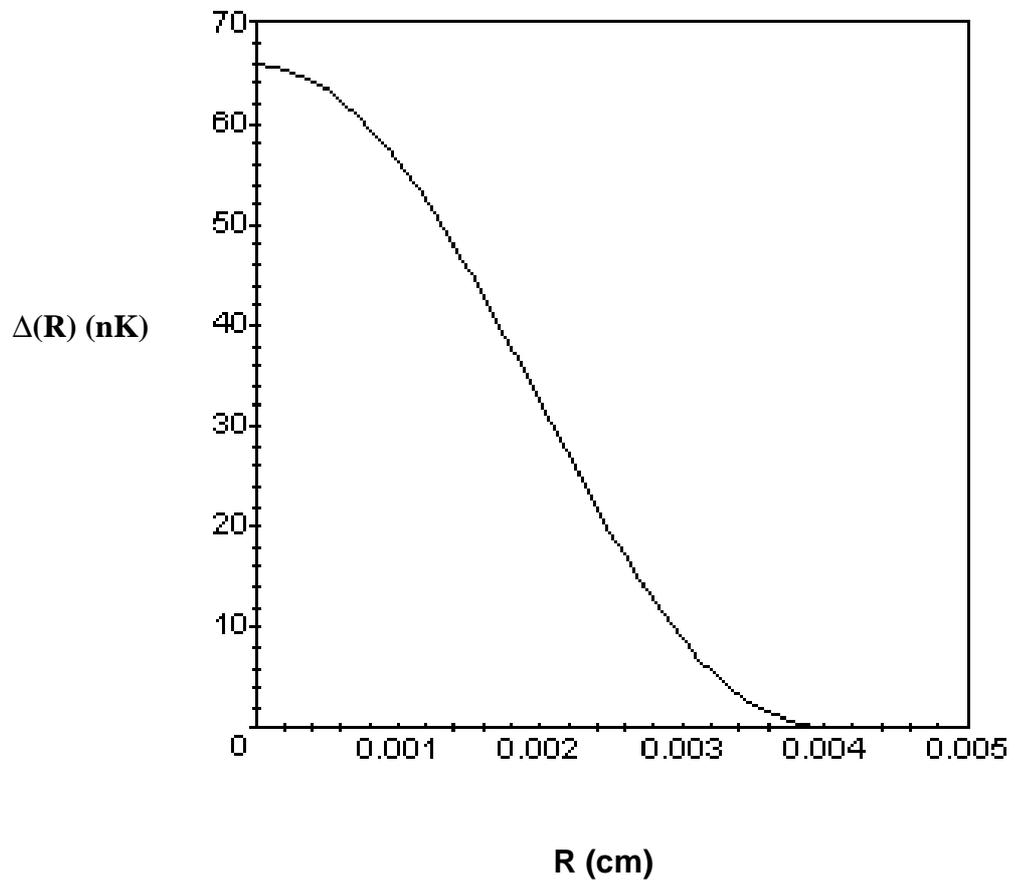

**Figure 2**

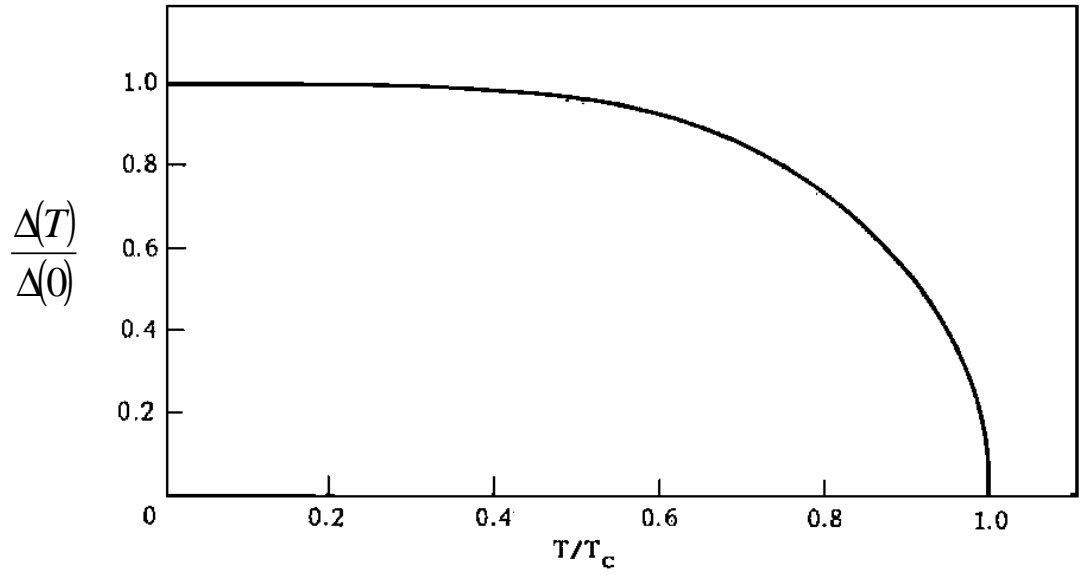

**Figure 3**

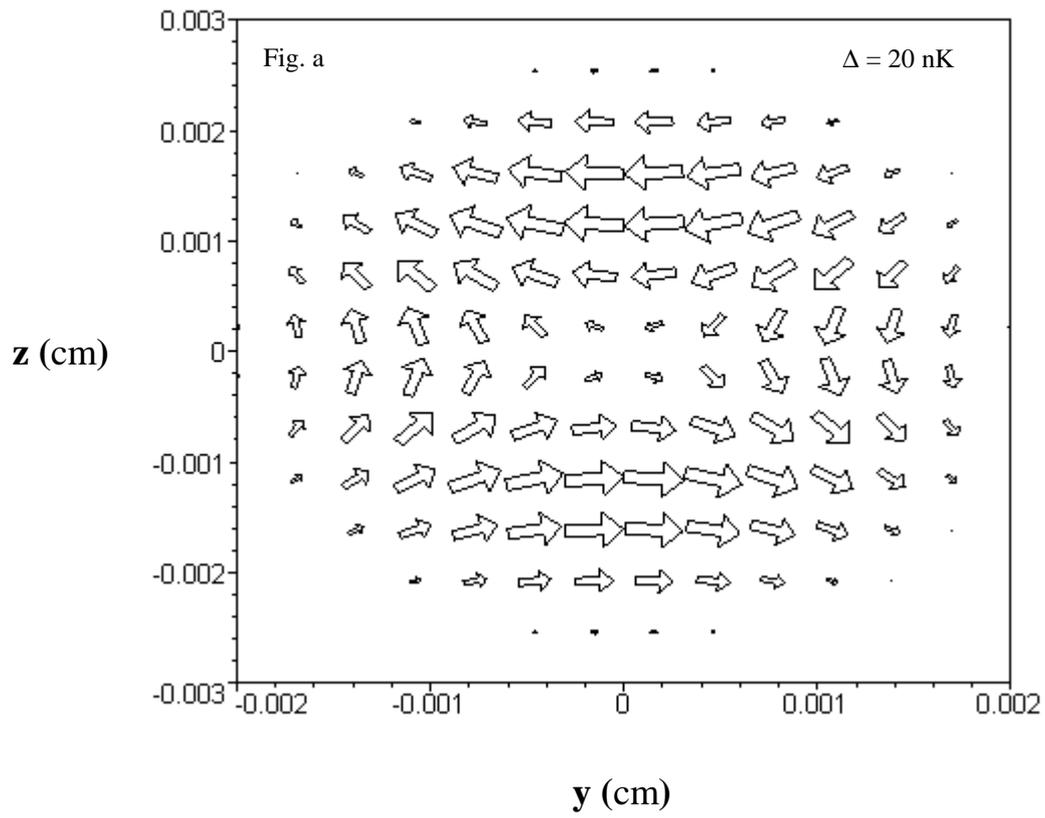

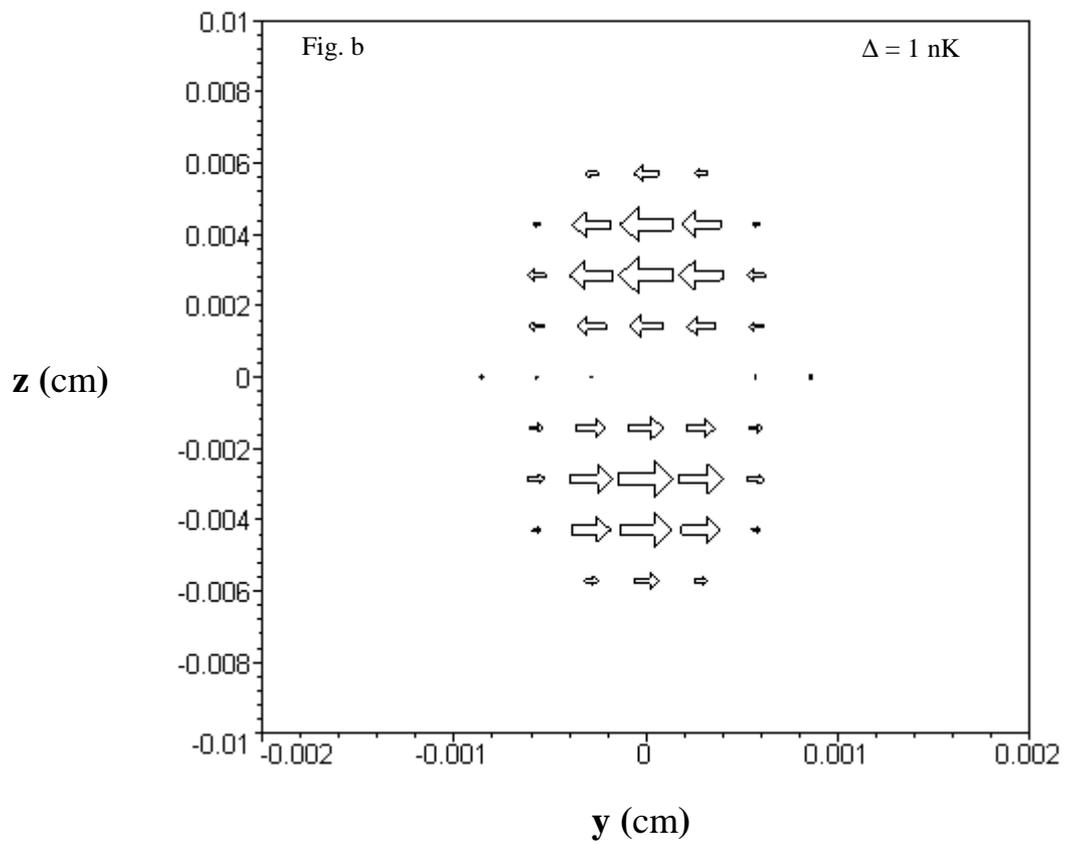

**Figures 4 a and b**

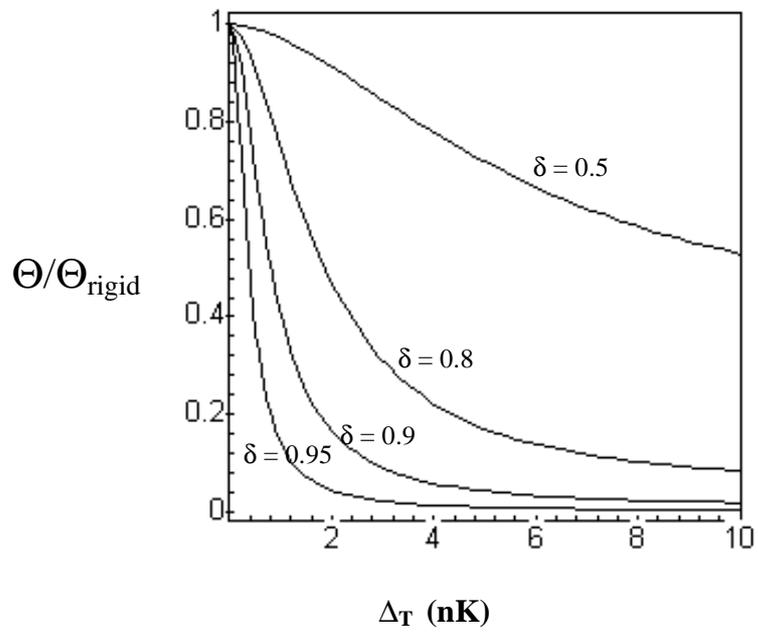

**Figure 5**